\documentclass[aps,prd,preprint,groupedaddress,showpacs,floatfix]{revtex4}
\usepackage{epsfig}
\usepackage{amsfonts}
\usepackage{float}
\usepackage{bm}
\usepackage{slashed}
\usepackage{url}

\begin{document}

%\preprint{}

\title{Beyond one-gluon exchange in the infrared limit of Yang-Mills theory}


\author{Marco Frasca}
\email[]{marcofrasca@mclink.it}
%\thanks{}
\affiliation{Via Erasmo Gattamelata, 3 \\ 00176 Roma (Italy)}


\date{\today}

\begin{abstract}
%We show that, in the infrared limit, Wilson loop gives a potential for a pure Yang-Mills theory that is in agreement with lattice expectations producing confinement. The potential is exactly given in this low-energy limit and the string tension is computed. Saturation of the potential at larger distances, or string breaking, is seen due to the mass gap and higher massive excitations of the gauge field, in agreement with lattice computations.
We analyze the Wilson loop for a pure Yang-Mills theory, using a decoupling solution in close agreement with lattice computations. At one-gluon exchange level it is seen that the potential cannot yield a linear rising contribution as expected for a confining theory. Next-to-leading order correction gives rise to a quartic term for momenta in the gluon propagator that, in agreement with Gribov's view, yields a linear confining term. This correction is due to a two-loop or sunrise integral that we need to evaluate in the low-momenta limit. In the infrared regime, the physical consistency of the theory is determined by a natural cut-off, arising from the integration of the classical equations of the theory, fixing in this way the regularization scheme.
\end{abstract}

%\keywords{}

\maketitle

\section{Introduction}

Since the groundbreaking work of Kenneth Wilson \cite{Wilson:1974sk}, a proof of confinement for quantum chromodynamics in the continuum limit is lacking. This is expected to be one of the major goals for our understanding of low-energy behavior of hadronic matter. Since then, a lot of fundamental work has been pursued reaching a deeper comprehension of the problem \cite{Greensite:2003xf,Greensite:2003bk,Greensite:2011zz,Bali:2000gf}.

Stated in its simplest way, the confinement problem can be expressed through the behavior of the expectation value of the Wilson loop that, for a free theory, is just reduced to the computation of a double path integral on the propagator of the theory. When this expectation takes the simple form $\exp(-\sigma_r S)$, being $S$ the area spanned by the loop and $\sigma_r$ a constant called string tension, one has that the {\sl area law} holds and the potential rises linearly with the distance granting completely bounded states. This result has been confirmed by computations on lattice \cite{Bali:2000gf,Greensite:2003xf} and is generally stated as ``Cornell potential'' being the sum of Coulombian and linearly rising terms.
%A linear increasing potential cannot be expected for all the range of distances. This has been shown clearly in lattice computations \cite{Kratochvila:2003zj,Buisseret:2006da} where the potential is seen to saturate at a certain distance and the string just breaks. This effect can be ascribed to the mass gap and higher excited states of the gauge field that acquire a mass and produce a screening effect.

The main problem to face to understand confinement is that the theory at low energies is strongly coupled. This means that perturbation theory is no more reliable to perform computations. Besides, very few techniques are available to manage the theory in this limit and, generally, one resorts to numerical studies on a lattice. Indeed, recent studies of the propagators of a pure Yang-Mills theory produced some striking results. These results on huge volumes, arriving to such a significant value as $(27fm)^4$, were presented at the Lattice 2007 Conference in Regensburg  \cite{Bogolubsky:2007ud,Cucchieri:2007md,Oliveira:2007px}. The shocking conclusion was that the scenario devised since then \cite{von Smekal:1997is,von Smekal:1997vx,Zwanziger:1991gz}, generally accepted as correct, was not describing the situation seen on the lattice in three and four dimensions: The gluon propagator was reaching a plateau at lower momenta with a finite non-zero value in zero, rather than being zero there, and the ghost propagator was behaving as that of a free massless particle while an enhanced one was expected. All in all, the running coupling was seen to bend clearly toward zero \cite{Bogolubsky:2009dc} without evidence of a non-trivial fixed point as was expected instead. Anyhow, there is not a widespread agreement on the definition of the running coupling in the infrared limit.

The scenario that emerged has an interesting interpretation. Assuming that an infrared trivial fixed point for a pure Yang-Mills theory exists \cite{Frasca:2007uz,Frasca:2009yp,Frasca:2008gi,Frasca:2010ce}, the theory can be easily managed through a perturbation expansion. The propagator we obtain can be used to get several properties of the theory and QCD as well \cite{Frasca:2011bd}. In this paper, we will show that this propagator, by itself, implies a screened potential and so, it needs a next-to-leading order correction to recover a Cornell potential and confinement. 
%This appears very easy as, with a trivial fixed point, we can apply results of a free theory for the Wilson loop. 
It is important to point out that, while Yang-Mills theory has an infrared trivial fixed point, this is no more true for QCD due to the presence of quarks. Besides, what makes small a running coupling is just the decreasing of momenta that are expected to go to zero like the fourth power. As we will see, the excellent agreement with numerical data, till a perfect coincidence for numerical solution of Dyson-Schwinger equations \cite{Aguilar:2004sw,Aguilar:2008xm}, is strongly supporting this view. This implies that the definition of the running coupling discussed in these papers is acceptable in the infrared limit and we are exploiting this possibility.

Finally, we emphasize that a decoupling solution for the gluon propagator does not grant a potential behaving as the Cornell potential with a linearly rising term but rather a screened potential. This has been recently shown, numerically solving Dyson-Schwinger equations, by Gonzalez, Vento and Mathieu \cite{Gonzalez:2011zc}. As we will show in this paper, the reason for this is that we need loop corrections to the propagator in the infrared limit. In this way, a quartic term in the momenta is seen to appear, in agreement with a Gribov-like propagator, providing the needed linearly rising term in the potential.

The paper is structured as follows. In sec.\ref{sec2}, we present the formalism applied to scalar field and Yang-Mills theories. In sec.\ref{sec3}, we show a comparison with numerical data strongly supporting the formalism. In sec.\ref{sec4}, we give the next-to-leading order correction to the propagator. In sec.\ref{sec5}, we discuss the renormalized gluon propagator we obtained. In sec.\ref{sec6}, we compute the interquark potential and the string tension. Finally, in sec.\ref{sec7}, we present our conclusions.

\section{Infrared quantum field theory\label{sec2}}

In the eighties, Carl Bender and others \cite{Bender:1978ew,Bender:1979et} proposed a new approach to cope with a strong coupled quantum field theory. This pioneering view was too radical and so, produced too singular results to be useful. But one can improve on it as we showed recently \cite{Frasca:2005sx,Frasca:2008gi,Frasca:2010ce}, assuming a bare coupling formally going to infinity. As we will show, immediate conclusions can be obtained both for scalar and Yang-Mills quantum field theories.

\subsection{Scalar field theory}

We consider a massless neutral scalar field with the following generating functional
\begin{equation}
    Z[j]=N\int [d\phi]\exp\left\{i\int d^4x\left[\frac{1}{2}(\partial\phi)^2-\frac{\lambda}{4}\phi^4+j\phi\right]\right\}.
\end{equation}
In order to get a strong coupling expansion, we formally rescale space-time coordinates as $x\rightarrow\sqrt{\lambda}x$. Then, the functional takes the form
\begin{equation}
    Z[j]=N\int [d\phi]\exp\left\{\frac{i}{\lambda}\int d^4x\left[\frac{1}{2}(\partial\phi)^2-\frac{1}{4}\phi^4+\frac{1}{\lambda}j\phi\right]\right\}.
\end{equation}
But $j$ is an arbitrary function and so we rescale it as $j/\lambda\rightarrow j$. Now, a perturbative expansion can be written down as
\begin{equation}
    \phi=\sum_{n=0}^\infty\lambda^{-n}\phi_n
\end{equation}
that yields the following terms into the action
\begin{eqnarray}
S_0&=&\int d^4x\left[\frac{1}{2}(\partial\phi_0)^2-\frac{1}{4}\phi_0^4+j\phi_0\right] \label{eq:ord0} \\
S_1&=&\int d^4x\left[\partial\phi_0\partial\phi_1-\phi_0^3\phi_1+j\phi_1\right] \label{eq:ord1} \\
S_2&=&\int d^4x\left[\frac{1}{2}(\partial\phi_1)^2-\frac{3}{2}\phi_0^2\phi_1^2
+\partial\phi_0\partial\phi_2-\phi_0^3\phi_2+j\phi_2\right]. \label{eq:ord2}
\end{eqnarray}
%From these, we are a step away from a proof of triviality of this theory in the infrared limit. Indeed, we just note that this functional demands, undoing the rescaling, to solve the equation
We note that this expansion requires that we are able to solve the classical equation of motion
\begin{equation}
\label{eq:phi0}
   \partial^2\phi_0+\lambda\phi_0^3=j
\end{equation}
in the limit $\lambda\rightarrow\infty$. Indeed, making use of this equation the following functional is obtained
\begin{equation}
   Z[j]\approx e^{i\int d^4x\left[\frac{1}{2}(\partial\phi_0)^2-\frac{\lambda}{4}\phi_0^4+j\phi_0\right]}
   \int[d\phi_1]
   e^{i\frac{1}{\lambda}\int d^4x\left[\frac{1}{2}(\partial\phi_1)^2-\frac{3}{2}\lambda\phi_0^2\phi_1^2\right]}.
\end{equation}
Here $\phi_0$ is a functional of the current $j$. In this form, we are not able to manage this expression. So, we make the Ansatz \cite{Frasca:2007id,Frasca:2007kb}
\begin{equation}
   \phi_0=\mu\int d^4x'\Delta(x-x')j(x')+\delta\phi
\end{equation}
and we need to solve the following equation for the Green function
\begin{equation}
    \partial^2\Delta(x-x')+\lambda[\Delta(x-x')]^3=\frac{1}{\mu}\delta^4(x-x').
\end{equation}
Here $\mu$ is an arbitrary constant with the dimension of energy arising in the solution of the classical theory. Then, using an iterative procedure, the next-to-leading order correction is
\begin{eqnarray}
\label{eq:delta}
    \delta\phi &=& \mu\lambda\int d^4x'\Delta(x-x')
    \left\{\mu\int d^4x''[\Delta(x'-x'')]^3j(x'')\right. \nonumber \\  
   &-&\left.\mu^3\left[\int d^4x''\Delta(x'-x'')j(x'')\right]^3\right\}+\ldots.
\end{eqnarray}
Higher order corrections can be similarly obtained. Finally, the propagator can be obtained explicitly \cite{Frasca:2005sx,Frasca:2006yx}. One has
\begin{equation}
\label{eq:prop}
    \Delta(p)=\sum_{n=0}^\infty\frac{B_n}{p^2-m_n^2+i\epsilon}
\end{equation}
being
\begin{equation}
    B_n=(2n+1)\frac{\pi^2}{K^2(i)}\frac{(-1)^{n}e^{-(n+\frac{1}{2})\pi}}{1+e^{-(2n+1)\pi}},
\end{equation}
with $K(i)=\int_0^{\frac{\pi}{2}}d\theta/\sqrt{1+\sin^2\theta}\approx 1.3111028777$ and a mass spectrum
\begin{equation}
\label{eq:ms}
    m_n = \left(n+\frac{1}{2}\right)\frac{\pi}{K(i)}\left(\frac{\lambda}{2}\right)^{\frac{1}{4}}\mu.
\end{equation}
%This can be easily seen as the Fourier transform in time at ${\bm p}=0$ of this propagator has the form $\langle\phi(0,t)\phi(0,t')\rangle = \sum_{n=0}^\infty C_n e^{-im_n(t-t')}$ showing that, in the infrared limit, the theory develops a mass gap. This is the key result of our analysis that also completes our proof of triviality of this scalar field theory in four dimensions in the infrared limit. 
We recognize that the theory develops a mass gap due to the finiteness of the coupling. This result would be very difficult to obtain with a weak coupling expansion. But, another striking result is that this theory is infrared trivial as the generating functional in this limit is just Gaussian
\begin{equation}
   Z_0[j]=N\exp\left[\frac{i}{2}\int d^4xd^4yj(x)\Delta(x-y)j(y)\right].
\end{equation}
%having the expected Gaussian form. We observe that the free excitation we have found in this limit entails a subset of excited states with the spectrum given by eq.(\ref{eq:ms}). 
%For the sake of completeness, we give here the next to leading order correction to this generating functional
This can be also evinced from the form of the propagator (\ref{eq:prop}) that presents just simple poles and can be ascribed to a free theory of massive particles with a superimposed spectrum of a harmonic oscillator. Finally, the next to leading order term takes the form
\begin{equation}
\label{eq:z1}
   Z[j]\approx Z_0[j]\int [d\phi]\exp\left\{\frac{i}{\lambda}\int d^4x\left[\frac{1}{2}(\partial\phi)^2-\frac{3}{2}\lambda\left(\int d^4x_1\Delta(x-x_1)j(x_1)\right)\phi^2\right]\right\}.
\end{equation}

\subsection{Yang-Mills theory \label{sec:ym}}

%Firstly, we review an approach devised in the eighties \cite{Goldman:1980ww,Cahill:1985mh}, where a current expansion was put forward, that gives a clear understanding of the stumbling block arisen in the studies of infrared Yang-Mills theory. Quantum Yang-Mills theory can be stated in its simplest form through the following generating integral
The foundations for a possible understanding of QCD at low-energies were posed on the eighties \cite{Goldman:1980ww,Cahill:1985mh}. People realized that, with a current expansion, a sensible low-energy limit was obtained. But not a great step forward was possible due to the stumbling block of an unknown gluon propagator. This was a major obstacle to complete this program. We will now see how a consistent scenarion can indeed be built on this basis. We have the generating functional
\begin{eqnarray}
\label{eq:zym}
    Z[j,\epsilon,\bar\epsilon]&=&N\exp\left\{-i\int d^4x\left[\frac{1}{4}{\rm Tr}F^2+\frac{1}{2\xi}(\partial\cdot A)^2
    +(\bar c^a\partial_\mu\partial^\mu c^a+g\bar c^a f^{abc}\partial_\mu A^{b\mu}c^c)\right]\right\}\times \nonumber \\
    &&\exp\left[i\int d^4x\left(j^a_\mu A^{\mu a}+\bar\epsilon^a c^a+\bar c^a\epsilon^a\right)\right]
\end{eqnarray}
being $F_{\mu\nu}^a=\partial_\mu A^a_\nu-\partial_\nu A_\mu^a+gf^{abc}A^b_\mu A^c_\nu$ the field strength and $A_\mu^a$ the vector potential. 
%Now, we can proceed as for the scalar field. In this case we just rescale $x\rightarrow\sqrt{N}gx$ being $N$ the number of colors. Then, in order to understand the behavior of Yang-Mills theory in the infrared limit, we need a way to manage the classical equations of motion (with standard notation)
Exactly in the same way as for the scalar field, we operate a rescaling of space-time variable through the 't Hooft coupling $\sqrt{Ng^2}$ for a SU(N) gauge group. So, again, we have to solve the classical equations of motion
\begin{equation}
\label{eq:noabe}
\partial^\mu\partial_\mu A^a_\nu-\left(1-\frac{1}{\xi}\right)\partial_\nu(\partial^\mu A^a_\mu) 
+gf^{abc}A^{b\mu}(\partial_\mu A^c_\nu-\partial_\nu A^c_\mu)
+gf^{abc}\partial^\mu(A^b_\mu A^c_\nu)
+g^2f^{abc}f^{cde}A^{b\mu}A^d_\mu A^e_\nu = -j^a_\nu.
\end{equation}
for an arbitrary gauge fixed through the parameter $\xi$. Now, we would like to study quantum field theory for Yang-Mills equations assuming that a trivial infrared fixed point exists. In order to get such a theory, we have to show that a set of classical solutions exist to build up the corresponding quantum field theory. Indeed, this is the case and such solutions are instantons \cite{Schafer:1996wv}. This can be proved quite easily by putting $A_\mu^a=\eta_\mu^a\phi(x)$ being $\eta_\mu^a$ some constant coefficients. Then, Yang-Mills equations collapse to
\begin{equation}
   \eta_\nu^a\partial^\mu\partial_\mu\phi-\left(1-\frac{1}{\xi}\right)\partial_\nu(\eta^a\cdot\partial\phi)+Ng^2\eta^a_\mu\phi^3=-j^a_\nu.
\end{equation}
In the Lorenz (Landau) gauge, this equation is further simplified and we can use the fact that $\eta_\nu^a\eta^{\nu a}=N^2-1$ and so we set $j_\phi=\eta_\nu^aj^{\nu a}$. In this case, the theory reduces to that of the scalar field and we can easily do quantum field theory for a trivial infrared fixed point. Quantum field theory will be build up on the instanton field $\phi$ that is replicated for each component of the Yang-Mills field.
Now, we are able to give explicitly the gluon propagator in the Landau gauge as
\begin{equation}
\label{eq:ymprop}   
  D_{\mu\nu}^{ab}(p)=\delta_{ab}\left(\eta_{\mu\nu}
  -\frac{p_\mu  p_\nu}{p^2}\right)\Delta(p)+O\left(\frac{1}{\sqrt{N}g}\right)
\end{equation}
with $\Delta(p)$ given by eq.(\ref{eq:prop}). Here $\lambda\rightarrow Ng^2$ and we interchange the constants $\mu$ and $\Lambda$. So, we can write the spectrum for SU(N) as
\begin{equation}
\label{eq:ms1}
    m_n = \left(n+\frac{1}{2}\right)\frac{\pi}{K(i)}\left(\frac{Ng^2}{2}\right)^{\frac{1}{4}}\Lambda.
\end{equation}  
This can be proven very easily from the definition of the two-point function. We have
\begin{equation}
   D_{\mu\nu}^{ab}(x-y)=\langle{\cal T}A_\mu^a(x)A_\nu^b(y)\rangle.
\end{equation}
Using our instanton solutions, one sees that
\begin{eqnarray}
   D_{\mu\nu}^{ab}(x-y)&=&\eta_\mu^a\eta_\nu^b\langle{\cal T}\phi(x)\phi(y)\rangle
   +O(1/\sqrt{N}g) \nonumber \\ 
   &=&\eta_\mu^a\eta_\nu^b\Delta(x-y)+O(1/\sqrt{N}g)
\end{eqnarray}
giving our result. %It is seen that the following Smilga's 
A possible choice for components of the gauge field 
%holds
is 
\begin{equation}
   \eta_\mu^a\eta_\nu^b=\delta_{ab}\left(\eta_{\mu\nu}-\frac{p_\mu p_\nu}{p^2}\right)
\end{equation}
being $\eta_{\mu\nu}$ the Minkowski metric. 
%We emphasize the striking conclusion that also the Yang-Mills field shares a trivial behavior in the infrared with the scalar field. The existence of this fixed point makes very easy to manage the theory in this limit.
We are now able to develop a quantum field theory for the Yang-Mills field supporting a trivial infrared fixed point. 

%For the ghost, we note that the mapping theorem grants its decoupling at the leading order. This can be seen immediately from the action of the Yang-Mills theory by direct substitution. So, at the infrared fixed point, the propagator of the ghost field is that of a free particle and we can write
For the sake of completeness, we give here the ghost propagator. Ghost field just decouples in the infrared limit due to the mapping theorem and so,
\begin{equation}
   G(p)=\frac{1}{p^2+i\epsilon}+O(1/\sqrt{N}g).
\end{equation}
Then, the generating functional for the gauge field in the infrared limit takes the following Gaussian form
\begin{eqnarray}
   Z[j]&=&N'\exp\left[\frac{i}{2}\int d^4xd^4y
   j^{\mu a}(x)D_{\mu\nu}^{ab}(x-y)j^{\nu b}(y)\right] \nonumber \\ 
   &+&O\left(\frac{1}{\sqrt{N}g}\right)
\end{eqnarray}
entailing triviality. 
%This is the limit we are interested in here. We note as the spectrum of the theory at the trivial infrared fixed point, given by eq.(\ref{eq:ms1}), is made of free massive excitations, with a superimposed spectrum of a harmonic oscillator, and so entails a mass gap. These can be considered as asymptotic states to start from to build up a perturbation theory in the infrared limit. This same conclusion does not hold in QCD as this theory has a non-trivial infrared fixed point due to the presence of quarks.
The spectrum has a mass gap with a superimposed spectrum of a harmonic oscillator as if such particles would have a structure. We note that infrared triviality does not apply to QCD due to the presence of quarks.

\section{Comparison with numerical data \label{sec3}}

%In this section we show how sound is the choice of the propagator describing low-energy physics starting from measurements obtained from lattice computations and numerical solution of Dyson-Schwinger equations. This kind of computations, relatively to the Landau gauge, span a two decade long period that has seen its main breakthrough on 2007 with the clear evidence that the gluon propagator in the Landau gauge is sitting on a plateau at very low-energy, reaching a finite non-zero value at zero momenta \cite{Bogolubsky:2007ud,Cucchieri:2007md,Oliveira:2007px} as also was proven in \cite{Aguilar:2004sw} for numerical Dyson-Schwinger equations. This means that, in order to show the soundness of our results given in the preceding sections, we have to compare with these computations.
A priori, it is not clear why our classical solutions should be selected to build up a quantum field theory. Even if our approach is consistent, we cannot be sure that something should enter, and it is instead neglected, into the solutions needed for the theory. So, we need to compare our conclusions with numerical data from lattice and numerical Dyson-Schwinger equations. These studies have been pursued for some decades and reached their maturity with a significant breakthrough on 2007 where the gluon propagator was seen to reach a finite non-zero value lowering momenta \cite{Bogolubsky:2007ud,Cucchieri:2007md,Oliveira:2007px}. This was also confirmed by numerical studies of Dyson-Schwinger equations \cite{Aguilar:2004sw}.

%We consider two kind of lattice computations: A set of volumes till $80^4$ directly obtained with measurements on the lattice and measurements at $128^4$ recovered from figure 2 in \cite{Cucchieri:2007md}. We are able to show in this way that, increasing the volume, our propagator describes even more accurately the one measured on the lattice in the deep infrared. We would like to point out that the mass gap is different for these two cases as it depends on the value of $\beta$. Then, using numerical Dyson-Schwinger results \cite{Aguilar:2008xm} where no volume problem arises, we see that our propagator perfectly matches the numerical solution in the deep infrared. Note that we consider a weak dependence on the gauge group as showed in \cite{Maas:2010qw} that is fully consistent with our discussion above.

\begin{figure}[H]
\begin{center}
\label{fig:rc}
\includegraphics[width=400pt,height=200pt]{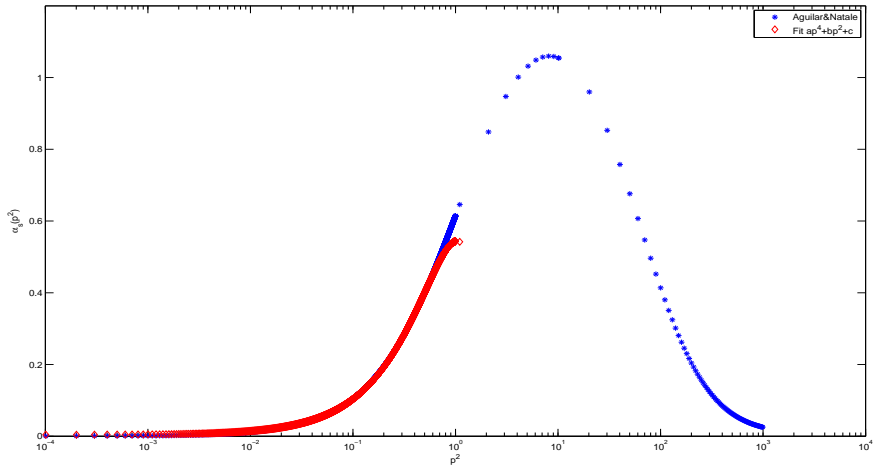}
\caption{Running coupling obtained by numerically solving Dyson-Schwinger equations (here $a=-0.5059\pm 0.0012$, $b=1.045\pm 0.001$ and $c=0.004\pm 0.001$.}
\end{center}
\end{figure}

%Added on 25 october to answer PRD
Firstly, the most important point to be assessed is the existence of the trivial fixed point at infrared (momenta going to zero). 
%This comes out from lattice computations as showed by Bogolubsky, Ilgenfritz, M\"uller-Preussker and Sternbeck \cite{Bogolubsky:2009dc}.
%\begin{figure}[H]
%\begin{center}
%\includegraphics[width=400pt,height=200pt]{Orlando.eps}
%\includegraphics{RunningCoupling.eps}
%\caption{Running coupling for $64^4$ and $80^4$ at $\beta=5.7$ taken from Ref.\cite{Bogolubsky:2009dc} (\URL{http://www.sciencedirect.com/science/journal/03702693}).\label{fig:stern}}
%\end{center}
%\end{figure}
%From fig.\ref{fig:stern} it is easily realized that the coupling reaches  trivial fixed points at higher and lower energies reaching a maximum in the intermediate regime. This is in perfect agreement with our scenario.
As already said, this can be seen from recent lattice computations \cite{Bogolubsky:2009dc} and numerical solution of Dyson-Schwinger equations \cite{Aguilar:2004sw,Aguilar:2008xm} provided one agrees on the definition of the running coupling at lower momenta. Assuming this, we will show below the behavior of the running coupling for the latter case (see fig. \ref{fig:rc}). We note also that, given a different definition of running coupling, a similar result has been given in \cite{Boucaud:2002fx}.

The definition of the running coupling is the one considered in \cite{Bogolubsky:2009dc} and proposed in \cite{Alkofer:2000wg}. It is given by
\begin{equation}
   \alpha_s(p^2)=\alpha_s(0) Z(p^2)H^2(p^2)
\end{equation}
having introduced the dressing functions for the gluon and ghost propagators respectively as $Z(p^2)=p^2\Delta(p^2)$ and $H(p^2)=p^2G(p^2)$. In our fit from numerical solution to Dyson-Schwinger equations, it is seen to be very near the zero value as momenta goes to zero. This result is identical to the one presented in \cite{Bogolubsky:2009dc}. We note that, with this definition, the running coupling goes to zero like $p^2$ while, with the definition given in \cite{Boucaud:2002fx}, it runs to zero like $p^4$. 

Below we give the comparison with numerical results for the propagators. Data for $80^4$ and Dyson-Schwinger equations were given directly from measurement while measurements at $128^4$ recovered from fig. 2 in \cite{Cucchieri:2007md}. One sees that, increasing volumes, our propagator tends to the measured one and the agreement improves neatly in the deep infrared. Different mass gaps are due to different values of $\beta$. For numerical Dyson-Schwinger equations \cite{Aguilar:2008xm}, one has no volume problem being already in the infinite limit and, indeed, we get perfect coincidence of our propagator with the measured one. Note that we consider a weak dependence on the gauge group as showed in \cite{Maas:2010qw} that is fully consistent with our discussion above.

\begin{figure}[H]
\begin{center}
\includegraphics{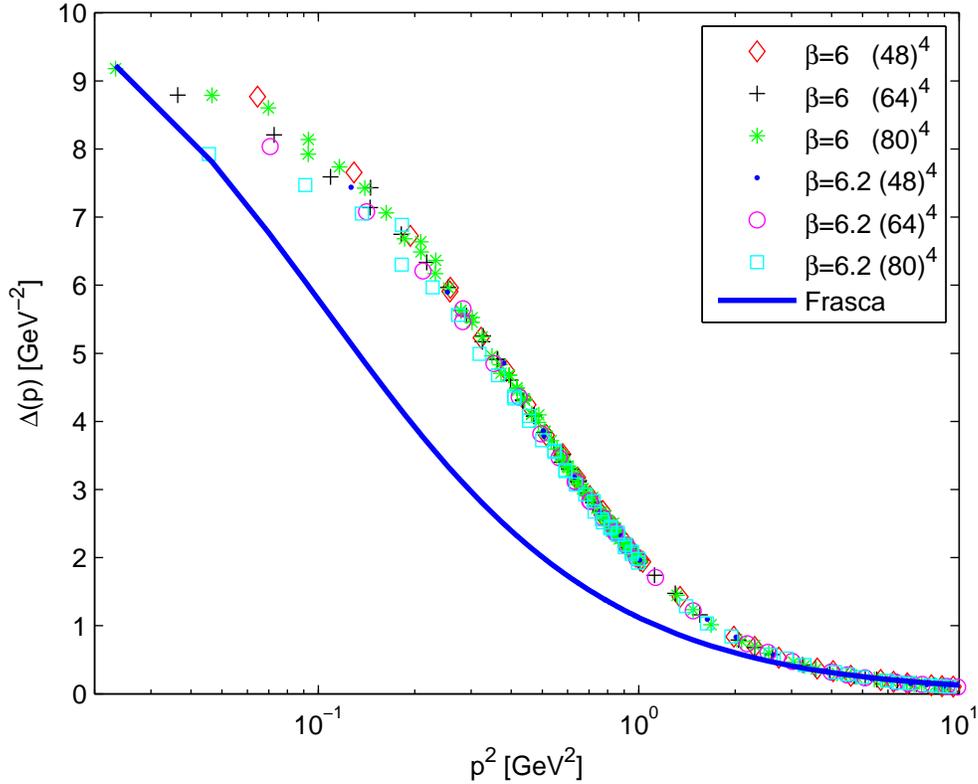}
\caption{Gluon propagator in the Landau gauge for SU(3), $\rm 80^4$ with a mass gap of $\rm m_0=321\ MeV$}
\end{center}
\end{figure}

\begin{figure}[H]
\begin{center}
\includegraphics{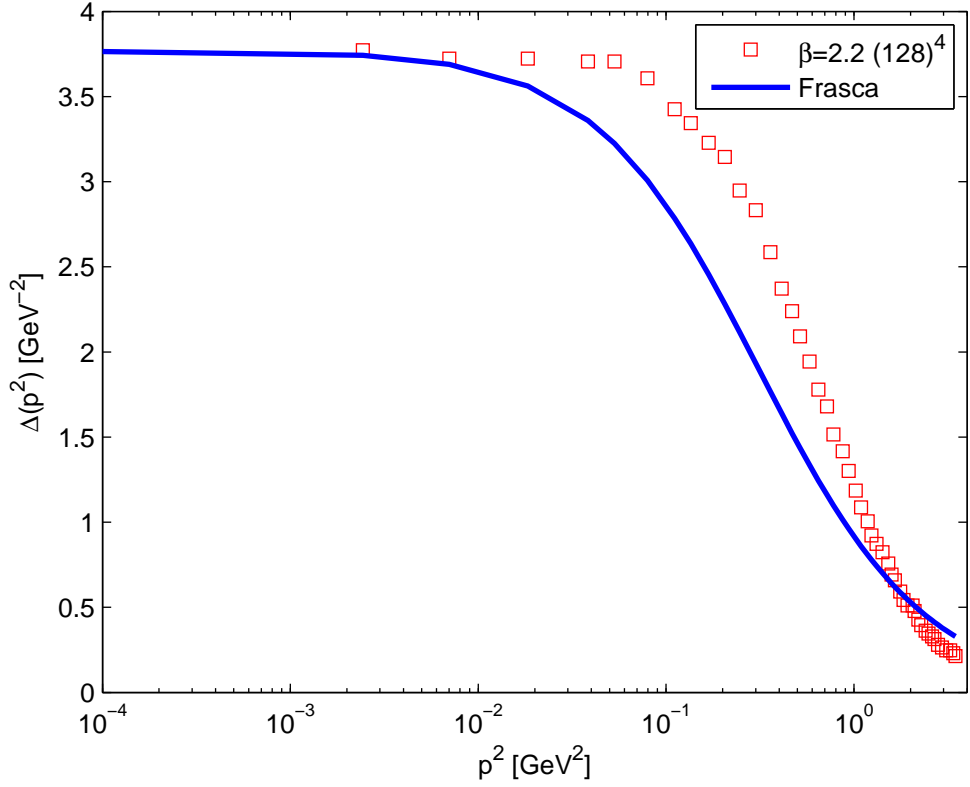}
\caption{Gluon propagator in the Landau gauge for SU(2), $\rm 128^4$ with a mass gap of $\rm m_0=555\ MeV$}
\end{center}
\end{figure}

\begin{figure}[H]
\begin{center}
\includegraphics[width=400pt,height=200pt]{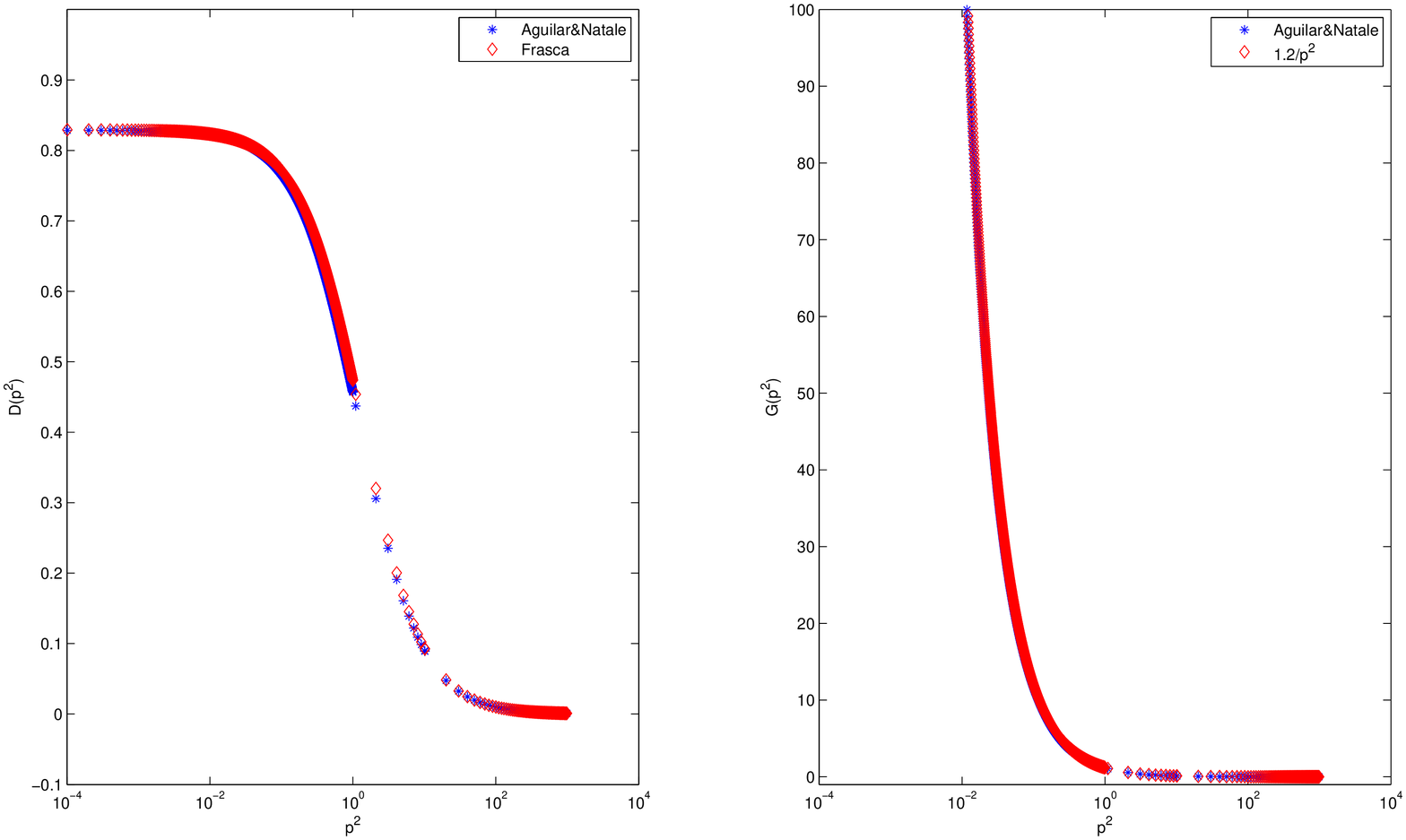}
\caption{Gluon and ghost propagators in the Landau gauge for SU(3) obtained by numerically solving Dyson-Schwinger equations and a mass gap $\rm m_0=399\ MeV$}
\end{center}
\end{figure}

%This agreement between lattice computations at increasing volume and the perfect match for the numerical Dyson-Schwinger equations with our propagator give a strong support to the idea that a pure Yang-Mills theory reaches an infrared trivial fixed point. This same conclusion cannot be drawn for QCD due to the presence of quarks.
These numerical results give a strong support to our scenario and to the conclusion of the existence of a trivial infrared fixed point for Yang-Mills theory.

\section{Two-loop correction \label{sec4}}

In sec. \ref{sec:ym}, we have shown that, for a Yang-Mills theory with a trivial infrared fixed point, it is enough to study the corresponding quantum theory of the instanton solutions represented by a massless scalar field with a quartic self-interaction. For our aims, we need just to evaluate the next-to-leading order correction to the propagator and derive the potential from it as shown in \cite{Gonzalez:2011zc}. This will move us away from the infrared fixed point and will provide a comprehension on the theory approaching such a fixed point.

So, from eq.(\ref{eq:delta}) we have to evaluate
\begin{equation}
    \Delta_R(x-x')-\Delta(x-x') = \mu^2\int d^4x''\Delta(x-x'')[\Delta(x''-x')]^3
\end{equation}
where $\Delta(x-x')$ is the one given in eq.(\ref{eq:prop}). We point out here that, for a Yang-Mills theory, we have to set $\lambda=Ng^2$. Turning to momentum space this gives
\begin{eqnarray}
\label{eq:gcorr}
   \Delta_R(p^2)-\Delta(p^2)&=&\lambda\frac{1}{\mu^2} \Delta(p^2)\int
   \frac{d^4p_1}{(2\pi)^4}\frac{d^4p_2}{(2\pi)^4}
   \sum_{n_1}\frac{B_{n_1}}{p_1^2-m_{n_1}^2}\times\\ \nonumber
   &&\sum_{n_2}\frac{B_{n_2}}{p_2^2-m_{n_2}^2}\sum_{n_3}\frac{B_{n_3}}{(p-p_1-p_2)^2-m_{n_3}^2}.
\end{eqnarray}
This integral is well-known in quantum field theory, it arises as a two loop sunrise diagram \cite{Caffo:1998du} and, in the form given above, has not a closed form value. But, we are in the infrared limit and we need its value just in the limit of small momenta. But before we take small momenta limit, consistently with our approximation technique, we need to take the limit $\lambda\rightarrow\infty$. This will give immediately
\begin{eqnarray}
%\label{eq:gcorr}
   \Delta_R(p^2)-\Delta(p^2)&=&\lambda\frac{1}{\mu^2} \Delta(p^2)
   \sum_{n_1,n_2,n_3}B_{n_1}B_{n_2}B_{n_3}
   \int\frac{d^4p_1}{(2\pi)^4}\frac{d^4p_2}{(2\pi)^4}\times \nonumber \\
   &&\left[\frac{1}{\lambda^\frac{3}{2}}\frac{1}{(2n_1+1)^2(2n_2+1)^2(2n_3+1)^2m_0^6}\right. \nonumber \\
   &-&\frac{1}{\lambda^2}\left(\frac{(p-p_1-p_2)^2}{(2n_1+1)^2(2n_2+1)^2(2n_3+1)^4m_0^8}\right. \nonumber \\
   &+&\left.\left.\frac{p_1^2}{(2n_1+1)^4(2n_2+1)^2(2n_3+1)^2m_0^8}+\frac{p_2^2}{(2n_1+1)^2(2n_2+1)^4(2n_3+1)^2m_0^8}\right)\right] \nonumber \\
   &+&O\left(\lambda^{-\frac{5}{2}}\right),
\end{eqnarray}
where we have set, using eq.(\ref{eq:ms}), $m_0=2^{-\frac{5}{4}}\mu\pi/K(i)\approx\mu$.
These integrals are really singular but we note that we cannot extend the integration range to infinity as we expect our computation to hold just for momenta $p<\mu$, the cut-off of the theory that emerged by integration of the classical equations. So, the theory has a natural cut-off. In this way we can immediately evaluate the correction with the following integrals:
\begin{eqnarray}
   \int\frac{d^4p_1}{(2\pi)^4}\frac{d^4p_2}{(2\pi)^4}&=&\frac{\mu^8}{\pi^8} \nonumber \\
   \int\frac{d^4p_1}{(2\pi)^4}\frac{d^4p_2}{(2\pi)^4}(p-p_1-p_2)^2&=&\frac{\mu^8}{\pi^8}\left(p^2+\frac{8}{3}\mu^2\right) \nonumber \\
   \int\frac{d^4p_1}{(2\pi)^4}\frac{d^4p_2}{(2\pi)^4}p^2_{1,2}&=&\frac{4}{3}\frac{\mu^{10}}{\pi^8}.
\end{eqnarray}
So, we finally get
\begin{equation}
%\label{eq:gcorr}
   \Delta_R(p^2)-\Delta(p^2)=\Delta(p^2)\left[\frac{1}{\lambda^\frac{1}{2}}\frac{27}{\pi^8}
   -\frac{1}{\lambda}\frac{3.3\cdot 48}{\pi^8}
   \left(1+\frac{3}{16}\frac{p^2}{\mu^2}\right)\right]+O\left(\lambda^{-\frac{3}{2}}\right).
\end{equation}
We note that the cut-off disappears everywhere except as a scale for the momenta contribution as it should. Now, we can interpret this as an expansion in inverse of $\sqrt{\lambda}$ of the denominator of the propagator and this means that we have to evaluate
\begin{eqnarray}
  (p^2-m^2_n)\left[1-\frac{1}{\lambda^\frac{1}{2}}\frac{27}{\pi^8}
   +\frac{1}{\lambda}\frac{3.3\cdot 48}{\pi^8}
   \left(1+\frac{3}{16}\frac{p^2}{\mu^2}\right)\right]&=& \nonumber \\
   p^2\left(1-\frac{1}{\lambda^\frac{1}{2}}\frac{27}{\pi^8}
   +\frac{1}{\lambda}\frac{3.3\cdot 48}{\pi^8}\right)
   +\frac{1}{\lambda}\frac{3.3\cdot 9}{\pi^8}\frac{p^4}{\mu^2}-m_n^2\left[1-\frac{1}{\lambda^\frac{1}{2}}\frac{27}{\pi^8}
   +\frac{1}{\lambda}\frac{3.3\cdot 48}{\pi^8}
   \left(1+\frac{3}{16}\frac{p^2}{\mu^2}\right)\right].
\end{eqnarray}
We get a mass renormalization factor causing the appearance of a $p^4$ term in agreement with Gribov expectations and, finally, looking at the mass term we see that
\begin{equation}
    m^2_n\left[1-\frac{1}{\lambda^\frac{1}{2}}\frac{27}{\pi^8}
   +\frac{1}{\lambda}\frac{3.3\cdot 48}{\pi^8}
   \left(1+\frac{3}{16}\frac{p^2}{\mu^2}\right)\right]=(2n+1)^2m_0^2\lambda^\frac{1}{2}\left[1-\frac{1}{\lambda^\frac{1}{2}}\frac{27}{\pi^8}
   +\frac{1}{\lambda}\frac{3.3\cdot 48}{\pi^8}
   \left(1+\frac{3}{16}\frac{p^2}{\mu^2}\right)\right]
\end{equation}
and we get a renormalized running coupling
\begin{equation}
   \lambda_R^\frac{1}{2}(p^2)=\lambda^\frac{1}{2}\left[1-\frac{1}{\lambda^\frac{1}{2}}\frac{27}{\pi^8}
   +\frac{1}{\lambda}\frac{3.3\cdot 48}{\pi^8}
   \left(1+\frac{3}{16}\frac{p^2}{\mu^2}\right)+O\left(\lambda^{-\frac{3}{2}}\right)\right].
\end{equation}
From this we get the renormalization constant for the field
\begin{equation}
   Z_\phi(p^2)=1-\frac{1}{\lambda^\frac{1}{2}}\frac{27}{\pi^8}
   +\frac{1}{\lambda}\frac{3.3\cdot 48}{\pi^8}
   \left(1+\frac{3}{16}\frac{p^2}{\mu^2}\right)+O\left(\lambda^{-\frac{3}{2}}\right).
\end{equation}

\section{Renormalized gluon propagator \label{sec5}}

Once we obtained the field renormalization constant $Z_\phi(p^2)$, we can provide a next-to-leading order formula for the gluon propagator. Let us introduce the constant $Z_0=Z_\phi(0)$ and $\lambda=Ng^2$. We can rewrite the gluon propagator in the Landau gauge in the following form
\begin{equation}
\label{eq:Dg}
   D_{\mu\nu}^{ab}(p^2)=\delta_{ab}\left(\eta_{\mu\nu}-\frac{p_\mu  p_\nu}{p^2}\right)
   \sum_{n=0}^\infty\frac{Z_0^{-1}B_n}{p^2+\frac{1}{\lambda}\frac{3.3\cdot 9}{\pi^8}\frac{p^4}{\mu^2}+m_n^2(p^2)}+O\left(\lambda^{-\frac{3}{2}}\right)
\end{equation}
where we have set
\begin{equation}
   m_n^2(p^2)=m_n^2\left[Z_0+\frac{1}{\lambda}\frac{3.3\cdot 9}{\pi^8}\frac{p^2}{\mu^2}+O\left(\lambda^{-\frac{3}{2}}\right)\right].
\end{equation}
We note that the coefficients $B_n$ are exponentially damped with $n$ and so, just the first few terms give the physics of infrared Yang-Mills theory. This means that our propagator agrees fairly well with the one obtained in the Refined Gribov-Zwanziger scenario \cite{Dudal:2003by,Dudal:2008sp,Dudal:2011gd}. So, we can attempt to evaluate the condensates after the correction is introduced. As already done in \cite{Frasca:2012vv}, we consider just two terms in our propagator and write down
\begin{equation}
   G(p^2)\approx \frac{Z_0^{-1}B_0}{p^2+\frac{1}{\lambda}\frac{3.3\cdot 9}{\pi^8}\frac{p^4}{\mu^2}+m_0^2(p^2)}+
   \frac{Z_0^{-1}B_1}{p^2+\frac{1}{\lambda}\frac{3.3\cdot 9}{\pi^8}\frac{p^4}{\mu^2}+m_1^2(p^2)}.
\end{equation}
In order to estimate the condensates, we note that $B_0+B_1\approx 1$ obtained from the exact relation $\sum_nB_n=1$ accounting for the exponential damping of higher coefficients. Now, by the definition given in \cite{Dudal:2010tf,Cucchieri:2011ig} of the condensate, we get
\begin{equation}
    \langle g^2A^2\rangle\approx-\frac{9}{13}\frac{N^2-1}{N}Z_0[(m_0^2+m_1^2)-(B_0m_1^2+B_1m_0^2)].
\end{equation}
In our previous work \cite{Frasca:2012vv} we get about $0.13\ GeV^2$ for this condensate and our result here is not so much different from it. Again, this is a proof of existence of such a condensate and our scenario agrees fairly well with that given in \cite{Dudal:2003by,Dudal:2008sp,Dudal:2011gd}.

\section{Wilson loop and potential \label{sec6}}

In order to compute the potential in a pure Yang-Mills theory at the infrared fixed point, we have to evaluate
\begin{equation}
    \left\langle {\rm tr}{\cal P}e^{ig\oint_{\cal C} dx^\mu T^aA^a_\mu(x)}\right\rangle = \frac{\int [dA][d\bar c][dc] e^{-\frac{i}{4}\int d^4x{\rm Tr}F^2+iS_g[\bar c, c]}
    {\rm tr}{\cal P}e^{ig\oint_{\cal C} dx^\mu T^aA^a_\mu(x)}}{\int [dA][d\bar c][dc] e^{-\frac{i}{4}\int d^4x{\rm Tr}F^2+iS_g[\bar c, c]}}
\end{equation}
being $S_g[\bar c, c]$ the contribution of the ghost field, $T_a$ the anti-hermitian generators of the gauge group and ${\cal P}$ the path ordering operator. In our case, in the infrared limit, we have a trivial fixed point. This implies that our generating functional takes a Gaussian form
\begin{equation}
     Z_0[j]=\exp\left[\frac{i}{2}\int d^4x'd^4x''j^{a\mu}(x')D_{\mu\nu}^{ab}(x'-x'')j^{b\nu}(x'')\right]
\end{equation}
and so, the evaluation of the Wilson loop is straightforwardly obtained as \cite{Pak:2009em}
\begin{equation}
     W[{\cal C}]=\exp\left[-\frac{g^2}{2}C_2(R)\oint_{\cal C}dx^\mu\oint_{\cal C}dy^\nu D_{\mu\nu}(x-y)\right]
\end{equation}
being $C_2(R)$ the quadratic Casimir operator that for SU(N) in the fundamental representation, $R=F$, is $C_2(F)=(N^2-1)/2N$. In our case, the propagator has the form given in sec.\ref{sec2} for the Landau gauge. The fall-off to large distances of this propagator grants that ordinary arguments to evaluate the integrals on the path apply. %The path is given in fig.\ref{fig:path}. 
%\begin{figure}[htb]
%\begin{center}
%\includegraphics{path.eps}
%\epsfig{file=path.eps,height=4in}
%\label{fig:path}
%\caption{Path for evaluation of Wilson loop.}
%\end{center}
%\end{figure}
Indeed, using Fourier transform one has
\begin{equation}
     W[{\cal C}]=\exp\left[-\frac{g^2}{2}C_2(R)\int\frac{d^4p}{(2\pi)^4}\Delta(p^2)\left(\eta_{\mu\nu}-\frac{p_\mu p_\nu}{p^2}\right)\oint_{\cal C}dx^\mu\oint_{\cal C}dy^\nu e^{-ip(x-y)}\right].
\end{equation}
We need to evaluate
\begin{equation}
     I({\cal C})=\eta_{\mu\nu}\oint_{\cal C}dx^\mu\oint_{\cal C}dy^\nu e^{-ip(x-y)}
\end{equation}
provided the contributions coming from taking into account the term $\frac{p_\mu p_\nu}{p^2}$ run faster to zero at large distances. This must be so also in view of the gauge invariance of Wilson loop. With the choice of time component in the loop going to infinity while distance is kept finite, we can evaluate the above integral in the form
\begin{equation}
     I({\cal C})\approx 2\pi T\delta(p_0)e^{-i{\bm p}\cdot{\bm x}}
\end{equation}
and we are left with
\begin{equation}
     W[{\cal C}]\approx \exp\left[-T\frac{g^2}{2}C_2(R)\int\frac{d^3p}{(2\pi)^3}\Delta({\bm p},0)e^{-i{\bm p}\cdot{\bm x}}\right]
\end{equation}
This yields
\begin{equation}
     W[{\cal C}]=\exp\left[-TV_{YM}(r)\right]
\end{equation}
being
\begin{equation}
     V_{YM}(r)=-\frac{g^2}{2}C_2(R)\int\frac{d^3p}{(2\pi)^3}\Delta({\bm p},0)e^{-i{\bm p}\cdot{\bm x}}.
\end{equation}
This equation should change if we consider a running coupling but, for the moment, we will not do this as does not change too much our conclusions. So, in our case we have to evaluate the integral \cite{Gonzalez:2011zc}
\begin{equation}
    V_{YM}(r)=-\frac{g^2}{4\pi R}C_2(R)\int_0^\infty dp p \Delta(p^2)\sin(pr)
\end{equation}
using eq.(\ref{eq:Dg}). If we neglect the contribution of the sunrise diagram we see that the propagator is a sum of Yukawa propagators. This means that the above integral will yield a sum of Yukawa potentials and we get an overall screened potential that does not grant confinement as discussed in \cite{Gonzalez:2011zc}. So, we see that in the deep infrared, one-gluon exchange does not grant a confining theory. But, accounting also for our two-loop correction, we will have to compute instead
\begin{equation}
      V_{YM}(r)=-\frac{g^2}{8\pi R}C_2(R)Z_0^{-1}\sum_{n=0}^\infty B_n\int_{-\infty}^\infty dp  
      \frac{p\sin(pr)}{p^2+\frac{1}{\lambda}\frac{3.3\cdot 9}{\pi^8}\frac{p^4}{\mu^2}+m_n^2(p^2)}.
\end{equation}
We can now evaluate this integral by rewriting it in the form
\begin{equation}
      V_{YM}(r)\approx-\frac{g^2}{8\pi R}C_2(R)Z_0^{-1}\frac{\pi^8\lambda\mu^2}{3.3\cdot 9}\sum_{n=0}^\infty B_n\int_{-\infty}^\infty dp  
      \frac{p\sin(pr)}{(p^2+\kappa^2)^2-\kappa^4}.
\end{equation}
where we have considered the limit $\lambda\rightarrow\infty$, $m_n^2,\ p^2=O(\sqrt{\lambda})$ even after a renormalization of mass $m_n^2\rightarrow Z_0m_n^2$ is taken, $\sum_nB_n=1$ and we have set $\kappa^2=\frac{\pi^8\lambda\mu^2}{3.3\cdot 9}$. This integral can be computed exactly to give
\begin{equation}
      V_{YM}(r)\approx-\frac{g^2}{8R}C_2(R)e^{-\frac{\kappa}{\sqrt{2}} r}\sinh(\frac{\kappa}{\sqrt{2}} r)
\end{equation}
that gives at small values of $\kappa R$
\begin{equation}
      V_{YM}(r)\approx -\frac{g^2}{8\pi r}C_2(R)\left[\frac{\pi}{\sqrt{2}}\kappa r-\frac{\pi}{2}\kappa^2r^2+O\left((\kappa r)^3\right)\right].
\end{equation}
So, remembering that $\lambda=d(R)g^2$, being $d(R)=N$ for SU(N) in the fundamental representation, we get a linear rising potential with $\sigma=\frac{\pi}{4}\frac{g^2}{4\pi}C_2(R)\kappa^2$ as expected. We can rewrite this as
\begin{equation}
    \sqrt{\sigma}\approx\frac{\pi^\frac{9}{2}}{2^\frac{7}{2}}g^2\sqrt{\frac{C_2(R)d(R)}{4\pi}}\mu.
\end{equation}
This result should be compared to the $d=2+1$ case \cite{Karabali:2009rg} that has
\begin{equation}
    \sqrt{\sigma_{d=2+1}}\approx g^2\sqrt{\frac{C_2(R)d(R)}{4\pi}}.
\end{equation}
The agreement is strikingly good in its functional form.

%We see that our linear approximation is not good for all distances but the potential is saturating due to the mass gap. This is in agreement with lattice evidence \cite{Kratochvila:2003zj,Buisseret:2006da}, as stated in the introduction.

\section{Conclusions \label{sec7}}

We have shown that, in the framework of our formalism, confinement is achieved for QCD satisfying the area law from the expectation value of the Wilson loop once next-to-leading order correction is accounted for. We started from the hypothesis of the existence of a trivial infrared fixed point for a Yang-Mills theory and this drove us to the fundamental conclusion that the scenario generally depicted by Wilson and Gribov is essentially correct.

As an aside, it interesting to note as our approach recovers in some limit the refined Gribov-Zwanziger scenario confirming the existence of a $A^2$ condensate. This condensate is relevant for the existence of a mass gap in the theory.

The overall agreement with lattice data and numerical studies of Dyson-Schwinger equations gives a serious hint that this view can represent an important track toward a full understanding of quantum field theory in the infrared limit.

\begin{acknowledgments}
I would like to thank Marco Ruggieri for useful comments and discussions and for providing code for numerical solving Dyson-Schwinger equations. I also thank Orlando Oliveira for sharing his numerical data with me. This problem was pointed out to me by Owe Philipsen at Bari Conference SM\&FT 2011. I would like to thank him for this.
\end{acknowledgments}

\end{document}